\documentclass{ws-procs9x6}

  \newcommand{\bq}{\begin{equation}}
  \newcommand{\eq}{\end{equation}}
  \newcommand{\bqa}{\begin{eqnarray}}
  \newcommand{\eqa}{\end{eqnarray}}
  
  \def\phv{\vec \phi}
  \def\gsim{\raise0.3ex\hbox{$>$\kern-0.75em\raise-1.1ex\hbox{$\sim$}}}

\begin{document}

\title{Critical behaviour and Scaling functions for the three-dimensional $O(6)$ spin model\\ with external field\footnote{\uppercase{T}his work is supported by the \uppercase{D}eutsche \uppercase{F}orschungsgemeinschaft under \uppercase{G}rant \uppercase{N}o. \uppercase{FOR} 339/1-2}}

\author{S. HOLTMANN and T. SCHULZE}

\address{Fakult\"at f\"ur Physik, Universit\"at Bielefeld, \\
D-33615 Bielefeld, Germany. \\ 
E-mail: holtmann, tschulze@physik.uni-bielefeld.de}

\maketitle

\abstracts{We numerically investigate the three-dimensional $O(6)$ model on 
$12^3$ to $120^3$ lattices. From Binder's cumulant at vanishing
magnetic field we obtain the critical coupling $J_c=1.42865(5)$ and verify this
value with the $\chi^2$-method. The universal value of Binder's cumulant at
this point is $g_r(J_c)=-1.94456(10)$. At the critical coupling we find
the critical exponents $\nu=0.818(5)$, $\beta=0.425(2)$ and $\gamma=1.604(6)$
from a finite size scaling analysis. We also determine the finite-size-scaling 
function on the critical line and the equation of state. Our $O(6)$-result for the 
equation of state is compared to the Ising, $O(2)$ and $O(4)$ results.}

\section{Introduction}
Our goal is the determination of universal quantities of the three-dimensional $O(6)$-invariant nonlinear $\sigma$-model, especially critical exponents and scaling functions. These quantities shall eventually be compared to those of 2-flavour staggered QCD with adjoint fermions. The relevant symmetry of this model is $SU(4)$, so it should be in the same universality class as the $O(6)$ spin model\cite{KL}.
The Hamiltonian of our model is defined as 
\bq
\beta\, {\mathcal H}\;=\;-J \,\sum_{<x,y>} \phv_x\cdot \phv_y \;-\; {\vec H}\cdot\,\sum_{x} \phv_x.
\eq
Here $x$ and $y$ are the nearest-neighbour sites on a three-dimensional  hypercubic lattice, $\phv_x$ is a $6$-component unit vector at site $x$  and $\vec H$ is the external magnetic field. The coupling constant $J$ is considered as inverse temperature, therefore $J=1/T$.
At $H=|{\vec H}|\neq 0$ the order parameter, the magnetisation $M$, is defined by
\bq
M\;=\;\langle\,\phi^{\parallel}\,\rangle,
\eq
where $\phi^{\parallel}$ is the longitudinal (parallel to $\vec
H$) component of $\phv$.
At $H=0$ this quantity vanishes for all couplings on the lattice, but we can use 
\bq
M\;=\; \langle \, |\phv|\,\rangle
\eq
as an approximate order parameter. We also measure the susceptibility $\chi$ and Binder's cumulant $g_r$, defined by
\bqa
\chi \;&=&\; V(\langle \: \phv^2\: \rangle \,-\,M^2),\\ 
g_r \;&=&\; \frac{\langle\:(\phv^2)^2 \: \rangle} {\langle \:\phv^2\:\rangle^2 }-3.
\eqa
We have simulated the model with a Wolff-Cluster-Algorithm,
using a ghostspin to emulate the external magnetic field. We used
hypercubic lattices with periodic boundaries and linear
extensions $L$ between 12 and 120. Near the critical coupling
$J_c$ we have made up to 200000 measurements per data point,
elsewhere about 20000 measurements. Between the
measurements we have used up to 400 cluster updates in the
broken phase and up to 1500 updates in the symmetric phase.

\section{Determination of the critical coupling}

\begin{figure}[t]
\setlength{\unitlength}{1cm}
\begin{picture}(13,4)
\epsfxsize=6.5cm
\put(-0.1,-2.4){
   \epsfbox{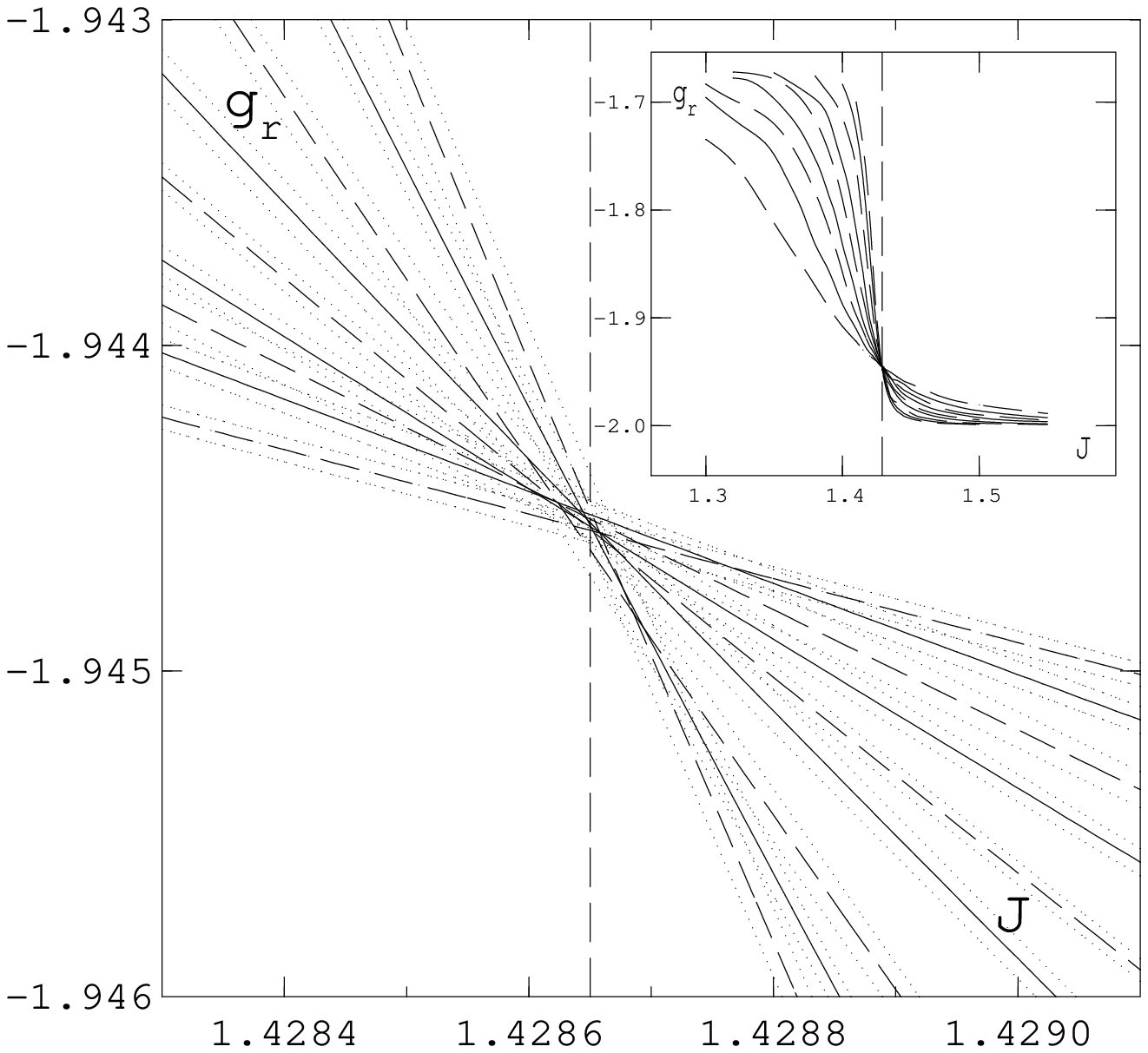}}
\epsfxsize=6.5cm
\put(5.5,-2.4){
   \epsfbox{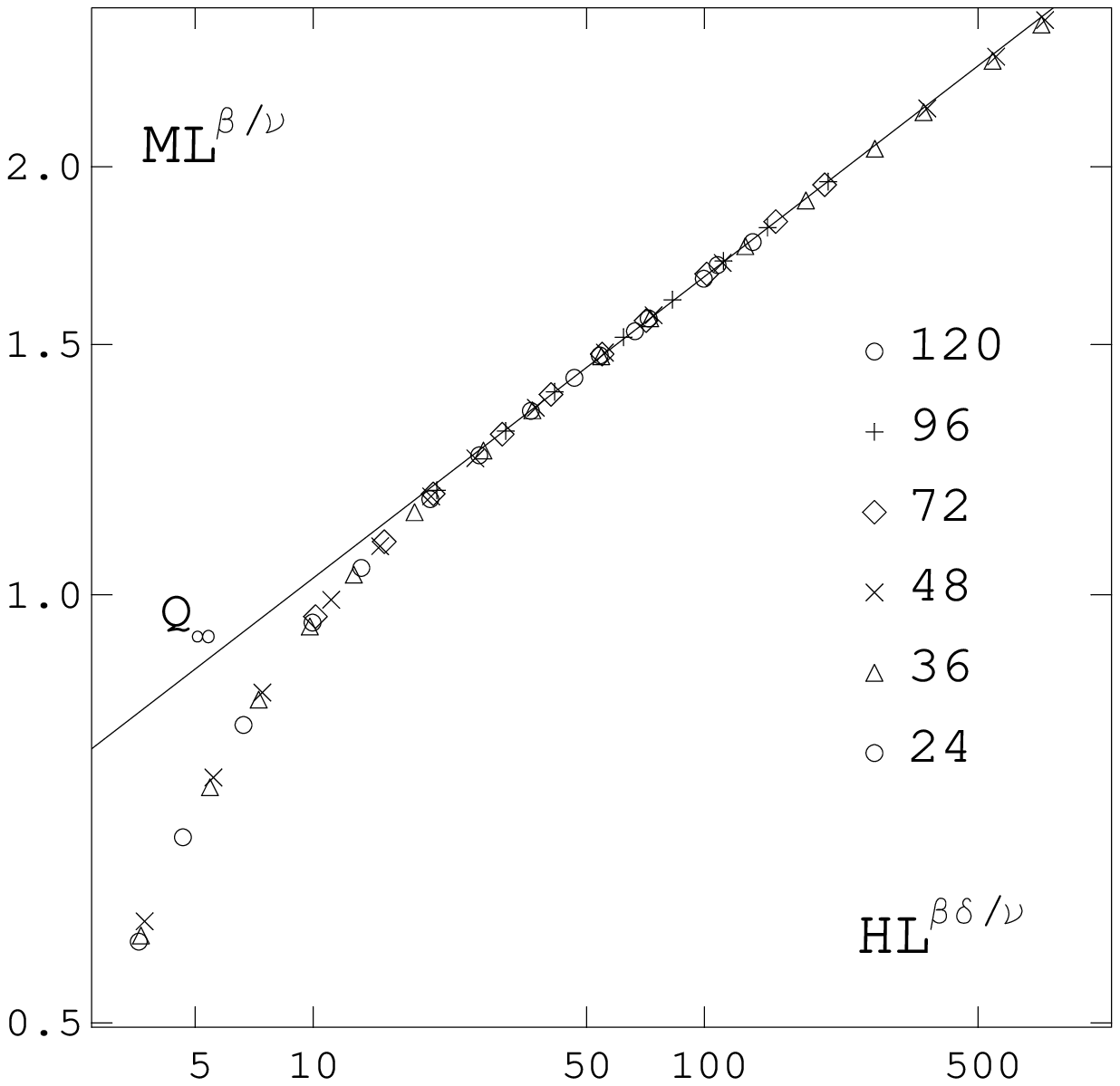}}
\end{picture}
\caption{The left plot shows the binder cumulant for the lattices with size $L=12$,$16$,$20$,$24$,$30$,$36$,$48$,$60$,$72$. The large lattices have a steeper slope then the small lattices. The right plot shows the finite-size-scaling function at the critical coupling.}
\label{binder}
\end{figure}

At $H=0$ the Binder cumulant can be described
by the finite-size-scaling function
\bq
g_r=Q_g(tL^{1/\nu};L^{-\omega }), 
\eq
where $t=\frac{T-T_c}{ T_c}$
is the reduced temperature and $\omega$ is the leading
irrelevant exponent. Therefore $g_r$ should be
independent of $L$ at the critical point $t=0$ apart from
corrections due to irrelevant scaling fields. The left plot in Fig.~\ref{binder} shows our reweighted data, with the dotted lines
representing the jackknife error corridors. Although he intersection points
coincide within the errors, there are some small corrections. As
the value of $\omega $ is unknown, we used Binder's
approximation\cite{Bi}
\bq
  \frac{1}{J_{ip}}\;=\;\frac{1}{J_c}\,+\,\frac{c_2}{\log b}
\eq
to extrapolate the intersection points $J_{ip}(L,b)$ of two
lattices with sizes L and $L^\prime =bL$. Fitting the results 
for the lattice sizes $L=12,16,20,24,30,36$ to a constant value we find
\bq
  \frac{1}{J_c}\;=\;0.699960(14)\;\;\Rightarrow \;\;J_c\;=\;1.42865(5).
\eq
We have checked this result with the $\chi^2$-method described 
in Ref. \refcite{E1}. From this value we derive the universal quantity 
$g_r(J_c)=-1.94456(10)$.\\
The result for $J_c$ is comparable with 
the result $J_c=1.42895(6)$ of Butera and Comi\cite{BC} using a high 
temperature expansion, but their errors seem to be underestimated.

\section{The critical exponents $\beta$, $\gamma$ and $\nu$}

For the determination of the critical exponents we use the
  following scaling relations at $t=H=0$:
\bqa
  M&\;=\;&L^{-{\beta /\nu}}\,(a_0\,+\,a_3L^{-\omega})\\  
  \chi &\;=\;&L^{\gamma / \nu}\,(b_0\,+\,b_3L^{-\omega})\\
  {\partial g_r\over\partial
  J}\;&=&\;L^{1/\nu}\,(d_0\,+\,d_3L^{-\omega}) 
\eqa
Fitting these relations to our data at lattices in the range $L=12-72$ we get 
$\beta /\nu =0.519(2)$ and $\gamma /\nu =1.961(3)$ from $M$ and $\chi$.
The error of these  quantities includes a
variation of $\omega $ between $0.5$ and $1.0$. From the
derivative of $g_r$ we get $1/\nu =1.223(5)$. Here we 
neglect the correction term, because it is zero within the errors. 
From these results we can calculate the critical
exponents $\beta =0.425(2)$, $\gamma =1.604(6)$ and $\nu =0.818(5)$ .
They fulfil the hyperscaling relations and are in complete agreement with the
results of Butera and Comi.\cite{BC}\\

\section{The equation of state}

\begin{figure}[t]
\setlength{\unitlength}{1cm}
\begin{picture}(13,4)
\epsfxsize=6.5cm
\put(-0.1,-2.4){
   \epsfbox{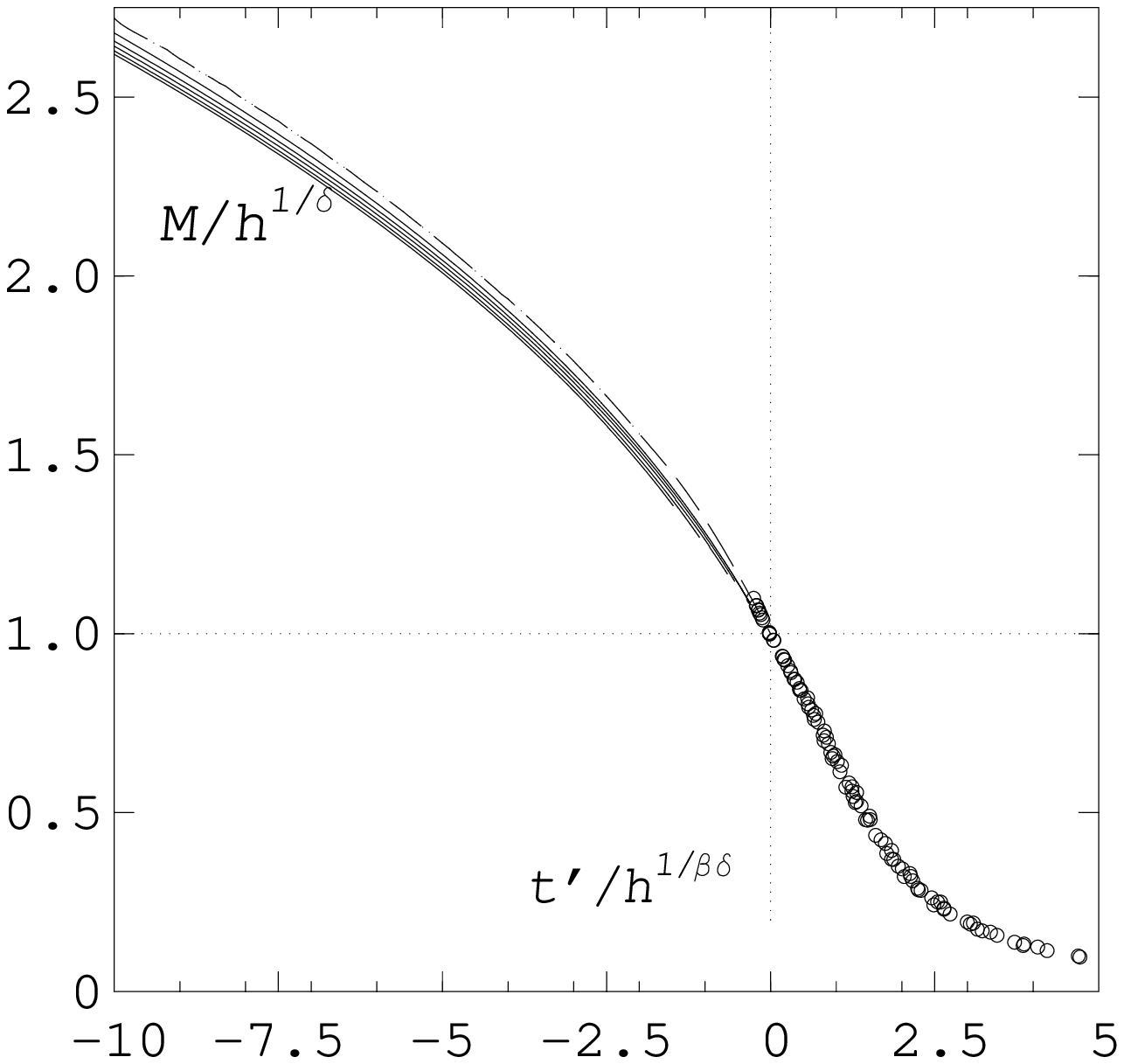}}
\epsfxsize=6.5cm
\put(5.5,-2.4){
   \epsfbox{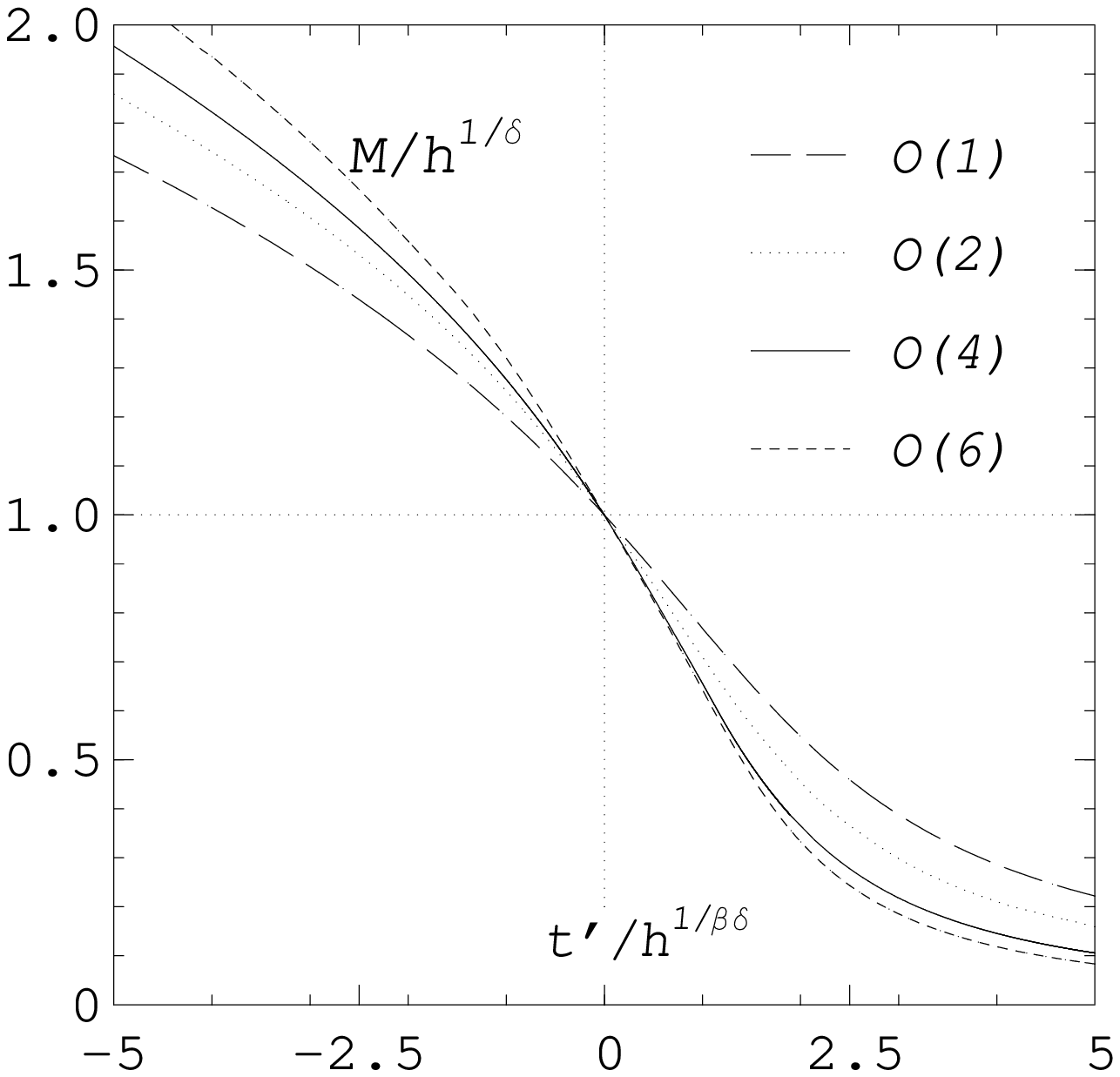}}
\end{picture}
\caption{The equation of state of $O(6)$ (left plot) and the comparison of the equations of state of $O(1)$ (Ising), $O(2)$, $O(4)$ and $O(6)$ (right plot).}
\label{skal}
\end{figure}

In the vicinity of $T_c$ the critical behaviour can be described
by the universal equation of state. It can be written in the form
\bq
M\;=\;h^{1/\delta} f_G(z'),\;\;\;\;z'\;=\;t'/h^{1/\beta\delta}
\eq
$t^\prime$ and $h$ are the normalized reduced temperature
$t^\prime =(T-T_c)/T_0$ and magnetic field $h=H/H_0$ with 
the normalization conditions $f_G(0)=1$ and
$f_G(z')=(-z')^\beta$ as $z'\rightarrow -\infty$. To get the
normalization 
constants $H_0$ and $T_0$ we have to determine the critical
amplitudes of the magnetisation on the critical isotherm and on
the coexistence line.
The infinite volume behaviour of $M$ is given by
\bq
M(T_c,H) \;=\;d_c H^{1/\delta}\,(1\,+\,d_c^1H^{\omega\nu_c})
\eq
including the leading correction term. From hyperscaling
relations we get for the exponents $\delta
=4.780(22)$ and $\nu_c = 0.4031(24)$. With this
ansatz, using only the data of the largest lattice at each value of $H$, we
get for the critical amplitude $d_c=0.642(1)$,
which translates to $H_0=d_c^{-\delta}=8.3(1)$.\\
In the broken phase the magnetisation is described by the ansatz
\bq
M(T<T_c,H)=M(T,0)+c_1(T)H^{1/2}+c_2(T)H.
\eq
The $H^{1/2}$ term is due to the Goldstone effect. As we have already 
seen with the $O(2)$\cite{E3} and $O(4)$\cite{E4} spin model, this ansatz fits 
very well and allows the extrapolation of data at nonzero $H$ to $H=0$. 
Fitting the results for $M(T,0)$ for several couplings ($J=1.45$,$1.47$,
$1.5$,$1.55$ and $1.6$) to the ansatz
\bq
M(T\le T_c,0)=B\,(T_c-T)^{\beta}\,[1\,+\,b_1\,(T_c-T)^{\omega
\nu}\,+\,b_2\,(T_c-T)]
\eq
we get $B=1.22(1)$, which leads to $T_0=B^{-1/\beta }=0.63(1)$.\\
The left plot in Fig.~\ref{skal} shows the scaling function $f_G(z')$. 
In the broken phase ($z'<0$) the solid lines are reweighted
data from different $J$-values. There are visible corrections,
so we use the ansatz
\bq
Mh^{-1/\delta}\;=\;f_G(z')\,+\,h^{\omega\nu_c}f_G^{(1)}(z')\,+
\,h^{2\omega\nu_c}f_G^{(2)}(z')
\eq
to obtain the universal value of $f_G(z')$ (dashed line). In
the symmetric phase the corrections are negligible.
In the right plot of Fig.~\ref{skal} we show a comparison of the $O(6)$ scaling
functions to those of some other $O(N)$ models, which have been
determined in Refs. \refcite{E2} (Ising), \refcite{E3} ($O(2)$) 
and \refcite{E4} ($O(4)$). As one
can see, the functions get steeper with increasing $N$.
These functions can eventually be used to determine the
universality class of staggered 2-flavour QCD both with
fundamental or with adjoint fermions. But until now QCD data has
too low statistics to distinguish between these scaling
functions, and also the used lattice sizes are too small in QCD to be
sure that one has reached the infinite volume limit.

\section{Finite-size-scaling function at $J_c$}

Because of the small lattice sizes used in QCD,
finite-size-scaling functions are a better tool for the comparison
of spin models to QCD. At $T_c$ and for small $H$ the different 
lattices should scale like 
\bq
M(T_c,H,L)\;=\;L^{-\beta / \nu}\,Q_M(HL^{\beta \delta /\nu})
\eq
where $Q_M$ is a universal scaling function. Our results are shown in
the right plot in Fig.~\ref{binder}. The solid line shows the
asymptotic behaviour in the limit $L\to\infty$
\bq
Q_M(z)\;=\;Q_\infty (z)\;=\;d_c\,z^{1/\delta},
\eq
which is observable for $z=HL^{\beta\delta /\nu}\gsim 40$.

\end{document}